# QoS Based User Driven Scheduler For Grid Environment

Sanjay Patel[1] and Madhuri Bhavsar[2]

[1]Department of Computer Engineering, SVBIT, Gandhinagar
`sanjaypatel54@gmail.com`
[2]Department of Computer Science & Engineering, Institute Of Technology,
Nirma University, Ahmedabad
`madhuri.bhavsar@nirmauni.ac.in`

## ABSTRACT

*As grids are in essence heterogeneous, dynamic, shared and distributed environments, managing these kinds of platforms efficiently is extremely complex. A promising scalable approach to deal with these intricacies is the design of self-managing of autonomic applications. Autonomic applications adapt their execution accordingly by considering knowledge about their own behaviour and environmental conditions.QoS based User Driven scheduling for grid that provides the self-optimizing ability in autonomic applications. Computational grids to provide a user to solve large scale problem by spreading a single large computation across multiple machines of physical location.*

*QoS based User Driven scheduler for grid also provides reliability of the grid systems and increase the performance of the grid to reducing the execution time of job by applying scheduling policies defined by the user. The main aim of this paper is to distribute the computational load among the available grid nodes and to developed a QoS based scheduling algorithm for grid and making grid more reliable.Grid computing system is different from conventional distributed computing systems by its focus on large scale resource sharing, where processors and communication have significant inuence on Grid computing reliability. Reliability capabilities initiated by end users from within applications they submit to the grid for execution. Reliability of infrastructure and management services that perform essential functions necessary for grid systems to operate, such as resource allocation and scheduling.*

## KEYWORDS

*Quality of Service (QoS); Scheduling; User Driven, Faiure To Repair Rate; Scheduling Instance*

## 1. INTRODUCTION

In most organizations, there are large amounts of underutilized computing resources.Most desktop machines are busy less than 5 percent of the time. In some organizations, even the server machines can often be relatively idle. A grid is built from multipurpose protocols and interfaces that address such issues like authentication, authorization and resource discovery. A grid allows its constituent resources to be used in a coordinated fashion to provide various qualities of service like response time, throughput etc. The User Driven Scheduler is intended to work as an resource managing module, queuing and scheduling of the Grid. The scheduler will offer managing batch jobs on Grid by scheduling CPU time according to user utility rather than system performance considerations. Autonomic applications adapt their execution accordingly by considering knowledge about their own behaviour and environmental conditions.QoS based User Driven scheduling for grid that provides the self-optimizing ability in autonomic applications. The main objective of this project is to provide the self-optimizing and QoS driven scheduling ability in autonomic applications desired by the user [1].

      Grid computing is an interesting research area that integrates geographically-distributed computing resources into a single powerful system. Many applications can benefit from such

integration. Examples are collaborative applications, remote visualization and the remote use of scientific instruments. Grid software supports such applications by addressing issues like resource allocation, fault tolerance, security, and heterogeneity. Parallel computing on geographically distributed resources, often called distributed supercomputing, is one important class of grid computing applications. Projects such as SETI@home, Intel's Philanthropic Peer-to-Peer Program for curing cancer and companies such as Entropia show that distributed supercomputing is both useful and feasible.

## 2. GRID SCHEDULING

In this work exploit the capabilities of Cellular Memetic Algorithms (cMAs) for obtaining efficient batch schedulers for Grid Systems. A careful design of the cMA methods and operators for the problem yielded to an efficient and robust implementation. Our experimental study, based on a known static benchmark for the problem, shows that this heuristic approach is able to deliver very high quality planning of jobs to Grid nodes and thus it can be used to design efficient dynamic schedulers for real Grid systems. Such dynamic schedulers can be obtained by running the cMAbased scheduler in batch mode for a very short time to schedule jobs arriving to the system since the last activation of the cMA scheduler [2].

## 3. EXIASTING ALGORITHMS AND METHODS (RELATED WORK)

### 3.1. Cellular Memetic Algorithm

In Memetic Algorithms (MAs) the population of individuals could be unstructured or structured. As in the case of other evolutionary algorithms, cMAs are high level algorithms whose description is independent of the problem being solved. In this work exploit the capabilities of Cellular Memetic Algorithms (cMAs) for obtaining efficient batch schedulers for Grid Systems. A careful design of the cMA methods and operators for the problem yielded to an efficient and robust implementation. Our experimental study, based on a known static benchmark for the problem,shows that this heuristic approach is able to deliver very high quality planning of jobs to Grid nodes and thus it can be used to design efficient dynamic schedulers for real Grid systems. Such dynamic schedulers can be obtained by running the cMAbased scheduler in batch mode for a very short time to schedule jobs arriving to the system since the last activation of the cMA scheduler[10].

### 3.2. Graph Theory

Services scheduling mainly acts according to some high-level application QoS parameters to carry on, for instance, the complete time, the reliability or the service cost and so on To develops a QoS aware Grid Services Scheduling optimal algorithm based on the complete time weight matrix. Grid Services Scheduling is a challenging problem under Open Grid Service Architecture (OGSA). A graph theory formal description is introduced into the Service Grid Model in this paper. The necessary and su_cient condition of complete matching of user job and service resources has been given and proved. Optimal Solution to matchmaking of grid jobs and grid services is developed based on the running time weight matrix, and the arithmetic has been verified by simulation analysis which proved to be more efficient than the alike arithmetic. The arithmetic has been implemented and running well[3].

### 3.3. Scheduling Instance

A scheduling instance is defined as a software entity that exhibits a standardized behaviour with respect to the interactions with other software entities The scheduling instance is the basic building block of a scalable, modular architecture for scheduling tasks, jobs, worflows, or applications in Grids[4].

### 3.4. Easy Grid AMS (Application Management System)

This section briey describes the EasyGrid AMS that is used in this work to manage the execution of a parallel MPI application on the computational grid. The EasyGrid AMS implements dynamic process creation and is automatically embedded into the MPI parallel application. It is not dependent on other grid system middleware, requiring only the Globus Toolkit and the LAM/MPI library to be installed[5].

## 4. EXIASTING METHODOLOGIES

A grid is created by installing Executors on each machine that is to be part of the grid and linking them to a central Manager component. The Windows installer setup that comes with the Alchemi distribution and minimal configuration makes it very easy to set up a grid.

### 4.1. Layered Architecture of Grid

Users can develop, execute and monitor grid applications using the .NET API and tools which are part of the Alchemi SDK. Alchemi offers a powerful grid thread programming model which makes it very easy to develop grid applications and a grid job model for grid-enabling legacy or non-.NET applications [1].

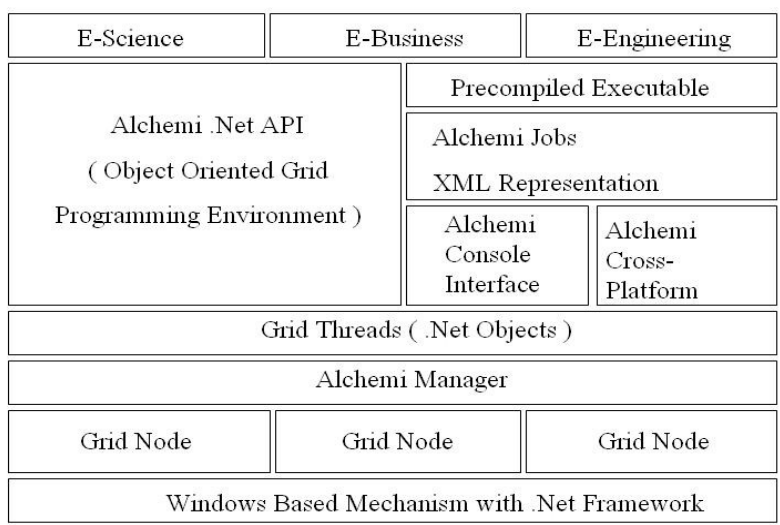

Figure 1: Layered Architecture of Grid

Alchemi layered architecture for a desktop grid computing environment is shown in Figure. Alchemi follows the master-worker parallel computing paradigm in which a central component dispatches independent units of parallel execution to workers and manages them. In Alchemi, this unit of parallel execution is termed grid thread and contains the instructions to be executed on a grid node, while the central component is termed Manager [1].

### 4.2 Existing Scheduling Mechanism

In existing scheduling mechanism all threads and computation done by system at kernel level. First of all application get divided in to different grid enabled process part and each small part called thread. When application divide in to different thread then system assign to each thread

Advanced Computing: An International Journal ( ACIJ ), Vol.2, No.1, January 2011different thread id. Each thread id(thread) divide for computation to available grid nodes. After computation each grid nodes return back thread or computation results to the head node [1].

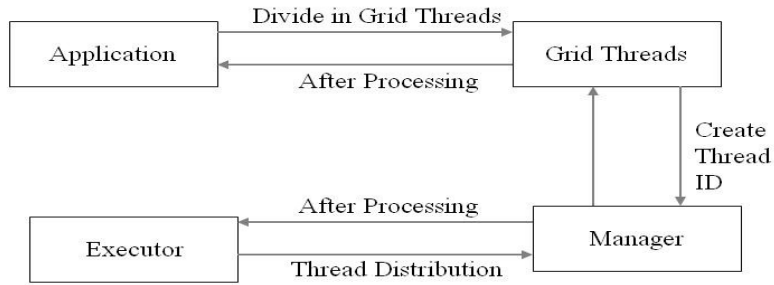

Figure 2: Existing Scheduling Mechanism

## 5. THE PROPOSED ALGORITHM AND METHODS

### 5.1 Process Sequence

There are four main entities in architecture, which are users, manager, schedulers and executor. A client is a user who submits a job to the system. A job refers to a collection of computation that the client wants to execute. The job is submitted by the client to the manager through a graphical user interface (GUI).

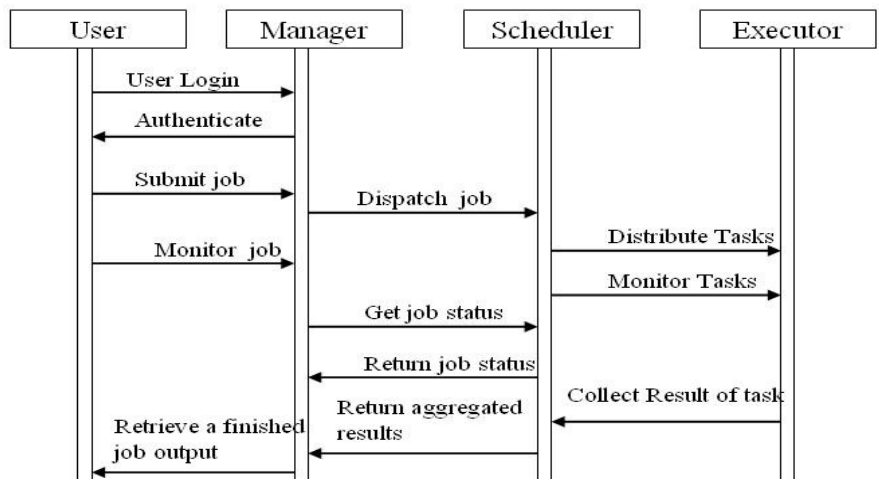

Figure 3: Job Processing Sequence

### 5.2 Design

A typical Grid consists of a number of services and a number of physical resources,including compute resources that are capable of hosting these services as well as storage resources, network resources etc. Grid applications are typically defined in terms of workows, consisting of one or more tasks that may communicate and cooperate to achieve their objective.

21

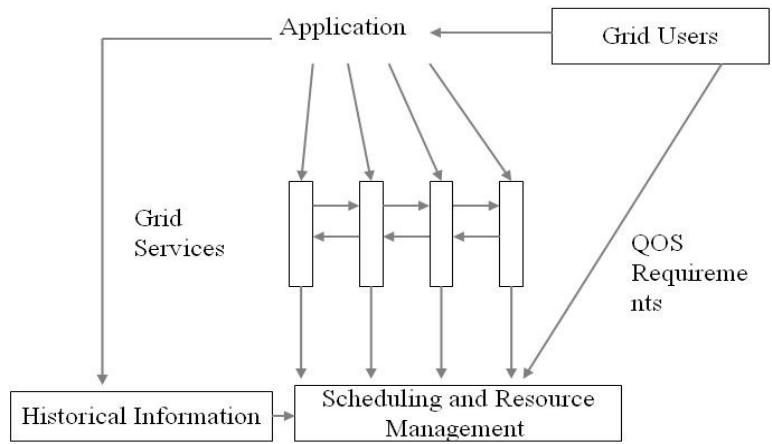

Figure 4: New Scheduling Mechanism

### 5.3 Algorithm Description

Failure of a resource while doing scheduling is not being considered at the time of allocating resources by the broker. Here, at the time of scheduling jobs, broker will consider only minimum cost of a resource along with MIPS of that resource. While doing scheduling, if resource fails to execute any job then such thing cannot be ignored when next time a job needs to be executed on that resource So for a resource a new parameter is added as failure rate which will consider success rate of a resource. If a resource is having 100% failure rate then that means that whenever a job is scheduled on that resource then it will surely fail to execute that job on that resource.

### 5.4 Flowchart

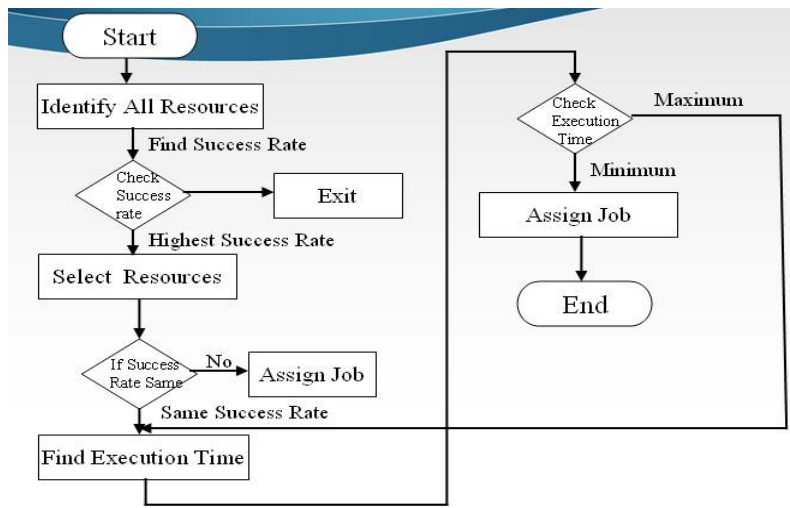

Figure 5: Flow Chart

In following figure given steps to identify best resource. First of all we identify all the resources in the grid. Then find success rate of each grid node.Compare success rate of each grid node and which grid node success rate is high then we select this resource for computation and also arrange grid node by success rate. but two grid node success rate is same then we also find execution time of each grid node. After find execution time of each grid node we compare execution time of grid node. We find minimum execution time of node that grid node we assign for computation. We rotate this steps.

## 6. USERS QOS REQUIREMENTS

- **Time:**
  Minimize execution time to increase the performance.
- **Reliability:**
  No. of failures for execution of workflows.
- **Fidebility:**
  Measurement related to the quality of the output of execution.

## 7. IMPLEMENT RELIABILITY IN GRID

Scientific applications have diverse performance and reliability requirements that are often difficult to satisfy, given the variability of underlying resources. Availability can vary due to failure of one or more critical services, load on one or more resource components, recovery from a failure, etc. Moreover, as Grid and web services continue to evolve, rapidly changing software stacks with concomitant configuration and service reliability challenges exacerbate application execution times and failures [13].

- **Reliability Programming Models**

    **1. Master Worker:** In the master-worker paradigm the master decomposes the problem into small tasks and distributes these tasks for execution.

    **2. Divide and Conquer :** The divide and conquer strategy partitions the problem into two or more smaller problems that can be solved independently and combined[14].

    **3. SPMD :** In the SPMD model, each task executes common code on different data. Failure of one task adversely affects the entire application, requiring global coordination.

## 8. RELIABILITY SPECIFICATION

In this section, we discuss the extensions required to the virtual grid description language to support reliability specifications. We define a high-level qualitative reliability metric space that can be used to request resources. The qualitative levels are mapped to well-defined quantitative reliability levels in the virtual grid to enable runtime monitoring and adaptation.

**Define a 5-point qualitative reliability scale that maps to quantitative levels of availability as follows: [11]**

**1) High Reliability ( 90-100% )**
**2) Good Reliability ( 80-89% )**
**3) Medium Reliability ( 70-79% )**
**4) Low Reliability ( 60-69% )**
**5) Poor Reliability ( 59-0% )**

## 9. PERFORMABILITY ANALYSIS

Grid systems are often able to survive the failure of one or more components and continue to provide service, but with reduced performance. Such behavior and status of systems with multiple interacting components is typically captured using stochastic process modeling [12]. The probability of staying in a certain state with respect to transition rates between states is used to quantify system performance and reliability. Markov Reward Models (MRM) are typically used to model gracefully degradable systems and capture joint performance and system reliability. A Markov reward model consists of a Markov chain that describes a systems possible states and an associated reward function.

## 10. RELIABILITY RESULTS

### 10.1 Without Reliability Model

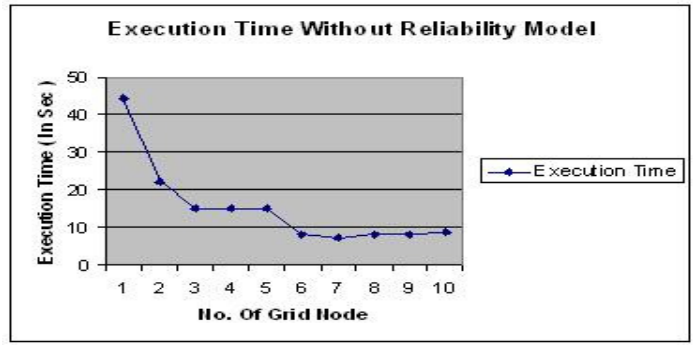

Figure 6: Without Reliability Model

### 10.2 With Reliability Model

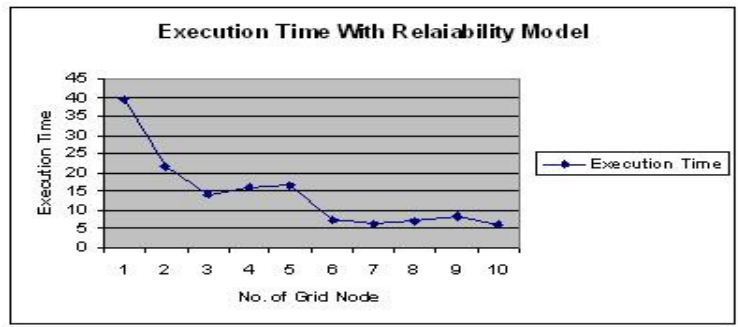

Figure 7: With Reliability Model

24

**10.3 Failure To Repair Rate**

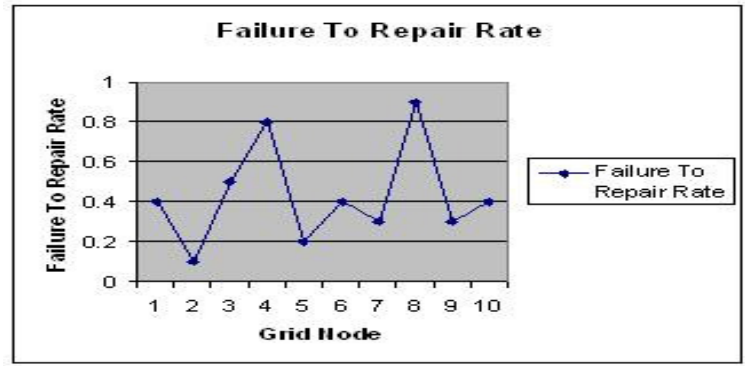

Figure 8: Failure To Repair Rate

## 11. Results Analysis

Showing the results of reliability we can say that using reliability model and programming method we can decrease the execution time of job and increase the performance of grid. Also find the failure to repair rate of each node and decide which node is better for our computation. We use values (from fig.8) for the failure-to-repair rate and performance degradation factors to study the variation in expected steady state reward rates. If we considered only performance, we would pick Grid node 2 as it completes the application most quickly. If we were to select a resource based on reliability, we would pick Grid node 8, the one with the lowest failure-to-repair ratio. Also we can say from fig.6 and 7 we run without reliability model program the execution time is high as compare to with reliability program and we run with reliability model program the execution time is low as compare to previous one. So we can say that reliability model is very important for decrease the execution time of job.

## 11. CONCLUSIONS

The objective of this paper is to deploy the computational grid by applying user driven scheduling policies with an improvement in QoS parameters. Major parameter considered in this paper is Time and Reliability.In Major Project,From the results we can say that using Qos based scheduler we can decrease the execution time and increase the performance of grid. QoS based user driven scheduler also provide large scale job by distribute across multiple grid nodes,and also reduce the execution time of a job and increase the performance of the grid. we can find failure to repair rate of each grid node.using reliability model and programming method we can decrease the execution time of job and increase the performance of grid. Also making grid more reliable. Showing the results of reliability(fig.6 and fig.7) we can conclude that using reliability model and programming method we can decrease the execution time of job and increase the reliable performance of grid.The objective of this paper is to deploy the computational grid by applying user driven scheduling policies with an improvement in QoS parameters.Major parameter considered in this paper is Time and Reliability.Also we can find failure to repair rate of each grid node.QoS based scheduling algorithm also provide large scale job by distributing across multiple grid nodes,and also reduce the execution time of a job and increase the performance of the grid.

## 12. ACKNOWLEDGEMENTS

I am also very thankful to faculty members and to my friends for their support. And at this moment, I would like to express my appreciation to my family members for their unlimited encouragement and support.

**Authors**

I am sanjay patel. I completed my M.Tech.(Computer science & engineering) from Nirma University in June – 2010. My Research area is Grid Computing. Currently I am working as a Assistant Professor in SVBIT.